\documentclass[reprint,prl,aps,twocolumn,showpacs,superscriptaddress,floatfix,footinbib,%bibliography=totoc
]{revtex4-2}
\usepackage{amsmath,amssymb}
\usepackage{hyperref}
\usepackage{color}
\usepackage{slashed}
\usepackage{placeins}

\usepackage{amsfonts}
\usepackage{graphicx, rotating}
\usepackage{epstopdf}
\usepackage{latexsym}
\usepackage{rotating}
\usepackage{braket}
\usepackage{url}
\usepackage{siunitx}
\usepackage[export]{adjustbox}
\usepackage{notes2bib}
\usepackage{graphicx}% Include figure files
\usepackage{dcolumn}% Align table columns on decimal point
\usepackage{bm}% bold math
\usepackage{xcolor}

\usepackage{tikz,xcolor,hyperref}
\definecolor{lime}{HTML}{A6CE39}
\DeclareRobustCommand{\orcidicon}{%
	\begin{tikzpicture}
	\draw[lime, fill=lime] (0,0) 
	circle [radius=0.16] 
	node[white] {{\fontfamily{qag}\selectfont \tiny ID}};	\draw[white, fill=white] (-0.0625,0.095) 
	circle [radius=0.007];	\end{tikzpicture}
	\hspace{-2mm}}
\foreach \x in {A,...,Z}{%
	\expandafter\xdef\csname orcid\x\endcsname{\noexpand\href{https://orcid.org/\csname orcidauthor\x\endcsname}{\noexpand\orcidicon}}
}
  \def\x{\xi}

\def\I{I}

\definecolor{mypurple}{rgb}{.5,0,.5}
\definecolor{myred}{rgb}{0.7, 0, 0}
\definecolor{myblue}{rgb}{0, 0, 0.7}
\definecolor{mygreen}{rgb}{0.04, 0.7, 0.5}
\definecolor{mygray}{rgb}{0.1, 0.1, 0.1}

\hypersetup{colorlinks,citecolor=myred,linkcolor=myblue,urlcolor=myblue,linktocpage=true}

\def\be   {\begin{equation}}   \def\ee   {\end{equation}}
\def\ba   {\begin{array}}      \def\ea   {\end{array}}
\def\bea  {\begin{eqnarray}}   \def\eea  {\end{eqnarray}}
\def\bean {\begin{eqnarray*}}  \def\eean {\end{eqnarray*}}

\def\bry{\begin{array}}
	\def\ery{\end{array}}

\def\GeV{\,{\rm GeV}}
\def\MeV{\,{\rm MeV}}

\def\eV{\,{\rm eV}}
\def\Mpl{ M_{\rm Pl}}

\newcommand{\skipnew}[1]{}

\def\Weizmann{\small{Department of Particle Physics and Astrophysics, Weizmann Institute of Science, Rehovot 761001, Israel}}
\def\WSU{\small{Department of Physics and Astronomy Wayne State University, Detroit, Michigan 48201, USA}}
\def\mainz{\small{Johannes Gutenberg-Universit\"at Mainz, 55128 Mainz, Germany}}
\def\ucb{\small{Department of Physics, University of California, Berkeley, California 94720, USA}}

\def\hhi{\small{Helmholtz-Institut, GSI Helmholtzzentrum f\"ur Schwerionenforschung, 55128 Mainz, Germany}}

\def\UD{\small{Department of Physics and Astronomy, University of Delaware, Newark, Delaware 19716, USA}}
\def\JQI{\small{Joint Quantum Institute, National Institute of Standards and Technology and the University of Maryland, Gaithersburg, Maryland 20742, USA}}
\def\ptb{\small{Physikalisch-Technische Bundesanstalt, Bundesallee 100, 38116 Braunschweig, Germany}}

\begin{document}

\date{\today}
\title{\Large\bfseries 
Oscillating nuclear charge radii as sensors for ultralight dark matter}

\author{Abhishek Banerjee\orcidA{}\, }
\email{abhishek.banerjee@weizmann.ac.il }
\affiliation{\Weizmann}

\author{Dmitry Budker\orcidB{}\,}
%\email{budker@uni-mainz.de}
\affiliation{\mainz}
\affiliation{\hhi}
\affiliation{\ucb}

\author{Melina Filzinger\orcidC{}\,}
%\email{melina.filzinger@ptb.de}
\affiliation{\ptb}

\author{Nils Huntemann\orcidD{}\,}
\email{nils.huntemann@ptb.de}
\affiliation{\ptb}

\author{Gil Paz\orcidE{}\,} 
%\email{gilpaz@wayne.edu}
\affiliation{\WSU}

\author{Gilad Perez\orcidF{}\,}
%\email{gilad.perez@weizmann.ac.il}
\affiliation{\Weizmann}

\author{Sergey Porsev\orcidG{}\,}
%\email{sporsev@gmail.com}
\affiliation{\UD}

\author{Marianna Safronova\orcidH{}\,}
%\email{msafrono@udel.edu}
\affiliation{\UD}
\affiliation{\JQI}

\begin{abstract}

We show that coupling of ultralight dark matter (UDM) to quarks and gluons would lead to an oscillation of the nuclear charge radius for both the quantum chromodynamic (QCD) axion and scalar dark matter, an effect which is of particular importance for heavy elements. Consequently, the resulting oscillation of electronic energy levels could be resolved with optical atomic clocks, and their comparisons can be used to investigate UDM nuclear couplings, which were previously only accessible with other platforms. We demonstrate this idea using the ${}^2S_{1/2} (F=0)\leftrightarrow {}^2F_{7/2} (F=3)$ electric octupole and ${}^2S_{1/2} (F=0)\leftrightarrow \,{}^2D_{3/2} (F=2)$ electric quadrupole transitions in \textsuperscript{171}Yb\textsuperscript{+}. Based on the derived sensitivity coefficients for these two transitions and a long-term comparison of their frequencies using a single trapped \textsuperscript{171}Yb\textsuperscript{+} ion, we find bounds on the scalar UDM-nuclear couplings and the QCD axion decay constant. These results are at a similar level compared to the tightest spectroscopic limits, and future investigations, also with other optical clocks, promise significant improvements.
\end{abstract}

\maketitle

Theories of ultralight dark matter (DM) bosons (scalar or pseudo-scalar) provide us with arguably the simplest explanation for the nature of this enigmatic substance. Ultralight DM (UDM) can be described as a classical field coherently oscillating with a frequency proportional to its mass $m_\phi$. 
Well-motivated models of UDM include the quantum chromodynamics (QCD) axion~\cite{Preskill:1982cy,Abbott:1982af,Dine:1982ah,Hook:2018dlk,DiLuzio:2020wdo}, the dilaton~\cite{Arvanitaki:2014faa}, the relaxion~\cite{Banerjee:2018xmn,Chatrchyan:2022dpy}, and possibly other forms of Higgs-portal models~\cite{Piazza:2010ye}. 
All of these predict that the UDM would couple to the Standard Model (SM) QCD sector, the quarks, and the gluons, leading to oscillations of nuclear parameters. Scalar UDM generically couples linearly to the hadron masses, whereas pseudoscalar UDM such as the QCD-axion couples quadratically to them, see e.g.~\cite{Kim:2022ype}. 
Optical clocks have been used to constrain DM couplings to electrons and photons (see~\cite{Antypas:2022asj} for a recent review). So far, limits on UDM nuclear couplings have been obtained via the $g$ factor dependence of hyperfine transition frequencies~\cite{Hees:2016gop,Kennedy:2020bac,Kobayashi2022,Flambaum:2002de,Zhang:2022ewz} and from molecular vibrations~\cite{Oswald:2021vtc}. In principle, they can also be derived from isotope mass shifts (via the reduced-mass dependence). However, the corresponding energy shifts scale as the inverse of the nuclear mass and therefore the sensitivity to the DM-nucleus coupling is suppressed. 

In this Letter, we propose and demonstrate using the oscillation of the nuclear charge radius for probing UDM-nuclear couplings with optical atomic clocks. This method is particularly effective for heavy atoms, opens complementary possibilities for investigating UDM-nuclear couplings, and increases the number of possible experimental platforms.

We derive the effects of nuclear charge-radius oscillations on electronic transitions and demonstrate the method using two optical clock transitions of $^{171}$Yb$^+$. 
Calculating the sensitivities of these transition frequencies to changes in the nuclear charge radius allows us to directly relate the QCD-axion and scalar UDM nuclear couplings to variations in the optical clock frequencies. From a 26-month optical atomic frequency comparison using a single $^{171}$Yb$^+$ ion, we obtain an experimental bound on UDM nuclear couplings.

The total electronic energy $E_\textrm{tot}$ of an atomic state contains the energies associated with the finite nucleus mass (mass shift, MS) and the non-zero nuclear charge radius $r_N$ (field shift, FS). They can be parameterized as~\cite{krane1991introductory}
\bea
\!\!\!\!\!\!\!
E_{\rm MS} \simeq K_{\rm MS} \frac{1}{m_A}\propto\frac{1}{A}\,\, {\rm and}\,\, E_{\rm FS} \simeq K_{\rm FS} \left<r_N^2\right>\propto A^{2/3}\,,
\label{eq:ms_fs_defn}
\eea
where $K_{\rm MS}$ and $K_{\rm FS}$ are the mass-shift and field-shift constants and the mass $m_A$ of an atom with atomic mass number $A$ is largely determined by the nuclear mass $m_N$. 
The variation of the total electronic energy associated with the nuclear degrees of freedom can be written as~\cite{king2013isotope}: 
\bea
\!\!\!\!\!\!\!
\left. \frac{\Delta E_{\rm tot}}{E_{\rm tot}}\right|_{\rm nuc} \simeq  - \frac{E_{\rm MS}}{E_{\rm tot}}\frac{\Delta m_N}{m_N}+\frac{E_{\rm FS}}{E_{\rm tot}}\frac{\Delta \left<r_N^2\right>}{\left<r_N^2\right>}\,.
\label{eq:deltaE_nul}
\eea
For heavy nuclei, the second term dominates, as in the case of \textsuperscript{171}Yb\textsuperscript{+} shown below. 
Thus, by comparing two electronic transition frequencies $\nu_a$ and $\nu_b$ of heavy atoms, we obtain 
\begin{align}
\frac{\Delta (\nu_a/\nu_b)}{(\nu_a/\nu_b)} = K_{a,b} \frac{\Delta \left<r_N^2\right>}{\left<r_N^2\right>}\,,\label{eq:d_nu_lim}
\end{align}
where we defined~\footnote{This discussion can be extended to transitions in two different atomic species when allowing for two different charge radii.}
\bea
    K_{a,b} \equiv \frac{K^{\nu_a}_{\rm FS} \,\langle r_N^2 \rangle}{\nu_a}
   -\frac{K^{\nu_b}_{\rm FS}\,\langle r_N^2 \rangle}{\nu_b} \,.
   \label{eq:K_ab}
\eea  

The mean squared nuclear charge radius of heavy elements is dominated by 
the distribution of protons within the nucleus rather than the charge structure of individual nucleons~\cite{Friar1975} (see also~\cite{Simonis:2017dny} and the references therein). 
The quantitative structure and details properties associated with the distribution of protons within the nucleus is related to the inter-nucleon interactions, which is  controlled by a variety of complex processes, such as the exchange of a single pseudo-scalar, scalar and vector mesons, at tree level, as well as double-exchange amplitude that corresponds to loop processes~\cite{Thomas:1981vc,Epelbaum:2008ga,Vanderhaeghen:2010nd,mandache_palade_2018,Ishii:2006ec,Ordonez:1995rz,Ordonez:1993tn}.
However, we are only interested in the log derivative of the charge radius with respect to the QCD fundamental parameters, $\Lambda_{\rm QCD}$, and the axion field (or the $\theta_{\rm QCD}$ parameter). More specifically, we only require to obtain the relevant scaling of the radius of heavy nuclei in terms of these fundamental parameters.  
To obtain the scaling we use mean-field computation, which has been shown to be able to capture the gross features of heavy nuclei (see {\it {e.g.}}~\cite{Reinhard:1989zi,Ring:1996qi,Gambhir:1990uyn} for reviews).
The relevant quantities are obtained by studying the dependence of $\left<r_N^2\right>$ on the QCD scale itself, and the pion mass square $m_\pi^2$, through its dependence on the QCD vacuum angle (see~\cite{Kim:2022ype} for instance) and/or the quark masses~\cite{Pich:1995bw}:
\begin{equation}
\frac{\Delta \left<r_N^2\right>}{\left<r_N^2\right>} \approx  \alpha \frac{\Delta\Lambda_{\rm QCD}}{\Lambda_{\rm QCD}}+ \beta\left.\frac{\Delta m_\pi^2}{m_\pi^2}\right|_{\Lambda_{\rm QCD}}\,.
\label{eq:rn_mpi_lqcd}
\end{equation}
 The calculation of $\alpha$ and $\beta$ is described in the Supplemental Material (and the full details in~\cite{toappear}), but in summary, we evaluate how the charge radius depends on the lowest resonance scalar coupling and mass, and the known dependence of these on $\Lambda_{\rm QCD}$ and the pion mass. Thus, it allows us to calculate $\alpha$ and $\beta$ in a theoretically controlled way. We find: 
 \begin{equation}
 \alpha =-1.1\,, \ \ \beta=-0.34\,,
 \end{equation}
and the errors of the mean field computation are estimated to be 30\% (see Supplemental Material for more detail).

As a first type of UDM, we consider a light scalar DM field, $\phi(t)$, interacting linearly with the up ($u$) and down ($d$) quarks and 
gluons ($G_{\mu\nu}$) as~\footnote{Our discussions can be extended for a more general ultra-light scalar DM couplings with the Standard Model (SM) QCD sector, and also to extend it to include the couplings to the quarks.}
\bea
\!\!\!\!\!\!\!\!\!\!
\mathcal{L} \supset -  \frac{\phi}{\sqrt{2}\Mpl}\left[\displaystyle
\sum_{q=u,d} d_{m_q}\,m_q \,\bar{q} q  + 
\frac{d_g\,\beta(g_s)}{2g_s} G^{\mu\nu}G_{\mu\nu}\right]\!\! ,
\eea
where $\beta(g_s)$ is the QCD beta function, $d_g$, $d_{m_q}$ are the coupling constants, $m_q$ is the mass of the quark $q$ and $\Mpl\simeq 2.4\times 10^{18}\GeV$ is the reduced Planck mass. We keep the color indices implicit.  
The oscillating DM background of the mass $m_\phi$, $\phi(t)= \sqrt{2\rho_{\rm DM}}/m_\phi\, \cos(m_\phi t)$, 
induces a small temporal component to $\alpha_s,$ and $\Lambda_{\rm QCD }$ and the quark masses as
\begin{eqnarray}
\!\!
\alpha_s(t) &=& \alpha_s(0)\left(1- 2 d_g\frac{\beta(g_s) \phi(t)}{g_s \sqrt{2}\Mpl} \right),\,
\frac{\partial\ln \Lambda_{\rm QCD}}{\partial \phi} = \frac{d_g}{\sqrt{2}\Mpl}\,\nonumber\\
\hat m(t) &=& \hat m(0)\left(1+ d_{\hat m}  \frac{\phi(t)}{\sqrt{2}\Mpl} \right),\,\frac{\partial\ln \hat m}{\partial \phi} = \frac{d_{\hat m}}{\sqrt{2}
\Mpl}\,,
\label{eq:phi_dg_dm}
\end{eqnarray}
where we define $\hat m = (m_u+m_d)/2$ and $d_{\hat m}= (m_u d_{m_u}+m_d d_{m_d})/(m_u+m_d)$. 
The variation of $m_\pi^2\propto \Lambda_{\rm QCD} \hat m$ \cite{Ubaldi:2008nf} for a fixed $\Lambda_{\rm QCD}$ can be related to $d_{\hat m}$ as 
\bea
\left.\frac{\Delta m_\pi^2}{m_\pi^2}\right|_{\Lambda_{\rm QCD}} &= &  d_{\hat m} \frac{\phi(t)}{\sqrt{2} \Mpl}\,.
\label{eq:mpi_phi}
\eea

Using Eqns.\,(\ref{eq:d_nu_lim},\ref{eq:K_ab},\ref{eq:rn_mpi_lqcd},\ref{eq:phi_dg_dm},\ref{eq:mpi_phi}), for a linearly coupled scalar DM of mass $m_\phi$, we obtain,
\bea
\!\!\!\!\!\!
\frac{\Delta (\nu_a/\nu_b)}{(\nu_a/\nu_b)} = K_{a,b}\, 
\Big[ 
\alpha\, d_g 
+ \beta\, d_{\hat m}
\Big]
\,\frac{\sqrt{2 \rho_{\rm DM}}}{m_\phi\,\sqrt{2}\Mpl}\,,
   \label{eq:scalar_DM_cr}
\eea
where we drop the explicit time dependence.

Let us now consider QCD axion models, where a pseudo-scalar field, the axion, $a$, couples to the gluon field,  
contributing a term to the Lagrangian density~\cite{Peccei:1977ur,Peccei:1977hh,Weinberg:1977ma,Wilczek:1977pj,Kim:1979if,Shifman:1979if,Zhitnitsky:1980tq,Dine:1981rt}:
$
\mathcal{L} \supset\frac{g_s^2}{32\pi^2}\frac{a}{f_a} G^{\mu\nu}\widetilde{G}_{\mu\nu}\,,
\label{eq:axion_gluon}
$
where $f_a$ is the axion decay constant, $g_s$ is the strong coupling constant, and $\widetilde G_{\mu\nu}$ is the dual gluon field strength.  
Considering interactions at energies much lower than the QCD confinement scale,
$\Lambda_{\rm QCD}$, this term gives rise to axion coupling to the hadrons. More specifically the pion mass depends on the axion as \cite{Ubaldi:2008nf,DiVecchia:1980yfw} 
\bea
\!\!\!\!\!\!\!\!\!\!\!\!\!\!\!
m_{\pi}^2(\theta_{\rm eff}) = \frac{\Lambda_{\rm QCD}^3}{f_{\pi}^2} \sqrt{m_u^2+m_d^2+2 m_u m_d \cos(\theta_{\rm eff})}\,,
\label{eq:m_pion}
\eea
where 
for the QCD-axion DM of mass $m_a$,
$\theta_{\rm eff}(t)= (a-\langle a\rangle)/f_a= \sqrt{2\rho_{\rm DM}}/(m_a f_a)\cos(m_a t)$\,.

The oscillating QCD axion DM induces an oscillating component to the pion mass at quadratic order as~\cite{Kim:2022ype}\footnote{As mentioned in~\cite{Ubaldi:2008nf,Kim:2022ype}, the nucleon mass also depends on the pion mass, so any variation in the pion mass would also lead to a variation in the nucleon mass as
\bea
\frac{\Delta m_{\mbox{\scriptsize{nucleon}}}}{m_{\mbox{\scriptsize{nucleon}}}} = 0.06\, \frac{\Delta m_\pi^2}{m_\pi^2}\,.
\label{eq:mn_mpi}
\eea
} 
\bea
\!\!\!\!\!\!\!\!\!\!\!\!
\frac{\Delta m_{\pi}^2}{m_\pi^2}=\frac{m_{\pi}^2(\theta_{\rm eff})-m_{\pi}^2(0)}{m_\pi^2(0)} \simeq  - \frac{m_u m_d\,\theta_{\rm eff}^2(t)}{2(m_u+m_d)^2}\,.
\label{eq:dmpi_dtheta}
\eea
Using Eqs.\,(\ref{eq:d_nu_lim},\ref{eq:K_ab},\ref{eq:rn_mpi_lqcd},\ref{eq:dmpi_dtheta}), we obtain, again without the explicit time dependence,
\bea
\frac{\Delta (\nu_a/\nu_b)}{(\nu_a/\nu_b)} = 
\, -\beta \,K_{a,b} \,\frac{m_u m_d}{(m_u+m_d)^2}\frac{\rho_{\rm DM}}{m_a^2f_a^2}\,.
   \label{eq:QCD_axion_cr}
\eea

The heavy \textsuperscript{171}Yb\textsuperscript{+} ion is a good candidate for the proposed search, as it features two optical clock transitions: the $(4f^{14}\,6s)\,^2\!S_{1/2} -\, (4f^{13}\,6s^2)\,^2\!F_{7/2} $ electric octupole ($E3$) and the $(4f^{14}\,6s)\,^2\!S_{1/2}-\,(4f^{14}\,5d)\,^2\!D_{3/2}$ electric quadrupole (E2) transition. We carried out isotope shift calculations for both of these transitions. 
According to our analysis, the MS is 30 times smaller than the FS for the $E3$ transition and 300 times smaller for the $E2$ transition. For this reason, we concentrate on the field shift in the following.

The FS operator, $H_\mathrm{FS}$~\cite{KozKor05}, 
modifies the Coulomb potential within the nucleus.
To find the FS coefficient $K_{\rm FS}$, we apply the finite-field method, adding $H_\mathrm{FS}$ to the initial Hamiltonian as a perturbation
with a coefficient $\lambda$:
$H \rightarrow H_\lambda = H + \lambda H_\mathrm{FS}$.
The coefficient $\lambda$ must be sufficiently large to make the effect of the field shift 
larger than the numerical uncertainty of the calculations, but small enough to keep the change in the energy linear in $\lambda$. In our calculation, we use $\lambda = \pm 0.01$.
Diagonalizing $H_\lambda$, we can find the eigenvalues $E_\lambda$ and determine $K_{\mathrm{FS}}$ as~\cite{KorKoz07,SafPorKoz18}:
\begin{equation}
 K_\mathrm{FS} = \frac{5}{6R^2} \frac{\partial E_\lambda}{\partial\lambda} \,,
\label{der}
\end{equation}
where $\partial \langle r_N \rangle / \langle r_N \rangle = \partial R /R \equiv \partial \lambda$, and we consider a nucleus as the uniformly charged ball with radius
$R = \sqrt{5/3}\, \langle r_N \rangle$.

The leading electron configurations of the $^2\!S_{1/2}$ and $^2\!D_{3/2}$ states have a filled $4f$ shell, while this is not the case in the $^2\!F_{7/2}$ state.
To calculate the energies of these three states, we use a 15-electron configuration interaction (CI) method, including the $4f$ shell
in the valence field.

We start from a solution of the Dirac-Hartree-Fock (DHF) equations by performing this procedure for the $[1s^2,...,4f^{14}6s]$ electrons. Then, all electrons are frozen and the electron from the $6s$ shell is moved to the $6p$ shell, and the $6p_{1/2,3/2}$ orbitals are constructed in the frozen core potential. All electrons are frozen again; the electron from the $6p$ shell is moved to the $5d$ shell, and the $5d_{3/2,5/2}$ orbitals are constructed. The remaining virtual orbitals are formed using a recurrent procedure described in~\cite{KozPorFla96,KozPorSaf15}.

In total, the basis set consists of five partial waves ($l \leq 4$) including orbitals up to $9s$, $9p$, $8d$, $8f$, and $7g$. 
The configuration space was formed by allowing single and double excitations for the even-parity states from the configurations $4f^{14} 6s$, $4f^{13} 6p5d$, and $4f^{13} 5d5f$
and for the odd-parity state from the configurations $4f^{14} 6p$, $4f^{13} 6s^2$, $4f^{13} 6p^2$, $4f^{13} 6s5d$, and $4f^{12} 6s^2 5f$.

To check the convergence of the CI method, we calculate the FS coefficients for four sets of configurations. First, we include
single and double excitations in the shells $6s$, $6p$, $5d$, $5f$, and $5g$ (we designate the set of excitations as [$6sp5dfg$]). Then we
sequentially included the single and double excitations to [$7sp6dfg$], [$8sp7dfg$], and [$9sp8dfg$].

In Table~\ref{FS} we present the FS coefficients $K_{\rm FS}$ found for the $^2\!S_{1/2}$, $^2\!D_{3/2}$, and $^2\!F_{7/2}$ states,
obtained for different sets of configurations. In the last column we list the FS coefficients $K^{\nu}_{\rm FS}$
determined for the transitions between the excited states $^2\!D_{3/2}$ and $^2\!F_{7/2}$ and the ground state, as
$K^{\nu}_{\rm FS} \equiv K_{\rm FS}(^2\!D_{3/2}, ^2\!F_{7/2}) - K_{\rm FS}(^2\!S_{1/2})$.
%================================================================================================================================================
\begin{table}[t]
\caption{\label{FS} The FS coefficients of levels $K_{\rm FS}$ and transitions $K^{\nu}_{\rm FS} \equiv K_{\rm FS}(^2\!D_{3/2}, ^2\!F_{7/2}) - K_{\rm FS}(^2\!S_{1/2})$ for various sets of basis configurations used in the calculation.}
\begin{ruledtabular}
\begin{tabular}{cccc}
  \multicolumn{1}{c}{Set of conf-s} & \multicolumn{1}{c}{Term}
& \multicolumn{1}{c}{$K_{\rm FS}$ } & \multicolumn{1}{c}{$K^{\nu}_{\rm FS}$}\\
\multicolumn{1}{c}{} & \multicolumn{1}{c}{}
& \multicolumn{1}{c}{$({\rm GHz}/{\rm fm}^2$)} & \multicolumn{1}{c}{$({\rm GHz}/{\rm fm}^2$)}\\
\hline \\[-0.6pc]
$[6sp5df\!g]$           & $^2\!S_{1/2}$   & -776.3   &        \\
                        & $^2\!D_{3/2}$   & -790.9   & -14.6  \\
                        & $^2\!F_{7/2}$ & -736.6   &  39.7  \\[0.5pc]

$[7sp6df\!g]$           & $^2\!S_{1/2}$   & -776.2   &        \\
                        & $^2\!D_{3/2}$   & -791.3   & -15.1  \\
                        & $^2\!F_{7/2}$ & -737.2   &  39.1  \\[0.5pc]

$[8sp7df\!g]$           & $^2\!S_{1/2}$   & -776.0   &        \\
                        & $^2\!D_{3/2}$   & -791.3   & -15.3  \\
                        & $^2\!F_{7/2}$ & -736.5   &  39.5  \\[0.5pc]

$[9sp8df\!g]$           & $^2\!S_{1/2}$   & -775.9   &        \\
                        & $^2\!D_{3/2}$   & -791.2   & -15.3  \\
                        & $^2\!F_{7/2}$ & -736.0   &  39.9  \\[0.5pc]
\hline \\[-0.6pc]
 Final                  & $^2\!S_{1/2} -\,^2\!D_{3/2}$ (E2)    &          & -15    \\
                        & $^2\!S_{1/2} -\,^2\!F_{7/2}$ (E3) &          &  40
\end{tabular}
\end{ruledtabular}
\end{table}
%================================================================================================================================================

As seen in Table~\ref{FS}, the coefficients $K^{\nu}_{\rm FS}$ are insensitive to increasing the number of configurations. 
To estimate a possible contribution from the core shells, we include six $5p$ electrons in the valence field and perform calculations in the framework of the 21-electron CI. The coefficients $K^{\nu}_{\rm FS}$ change
only at the level of 2\%. Assuming that the contribution from other core shells can be as large as 10\% and also taking
into account a possible contribution from valence-valence correlations beyond the $[9sp8df\!g]$ set of configurations,
we estimate the uncertainty of $K^{\nu}_{\rm FS}$ at the level of 12-15\%. 
Using the final values given in Table~\ref{FS}, we find that the ratio of the FS coefficients $K_{\rm FS}$ for the $E3$ and $E2$ transitions is $-2.7(6)$. This result agrees well with that obtained in a recent experimental determination of high precision $K^{\nu_{\textrm{E3}}}_{\rm FS}/K^{\nu_{\textrm{E2}}}_{\rm FS} = -2.1962536(14)$~\cite{HurCraCou22}.

The frequencies of the investigated E3 and E2 transitions are $\nu_{\textrm{E3}} \approx 6.42\times10^{14}\,{\rm Hz}$ and
$\nu_{\textrm{E2}} \approx 6.88\times10^{14}\,{\rm Hz}$, respectively. Using the calculated FS coefficients $K^{\nu_{\textrm{E2}}}_{\rm FS} = -15\,{\rm GHz}/{\rm fm}^2$ and $K^{\nu_{\textrm{E3}}}_{\rm FS} = 40\,{\rm GHz}/{\rm fm}^2$ and $\langle r_N \rangle \approx 5.3\,{\rm fm}$~\cite{marinova_angeli}, we obtain
\bea
\!\!
K_{\rm {E3,E2}}= \left(\frac{K^{\nu_{\textrm{E3}}}_{\rm FS}}{\nu_{\textrm{E3}}}
  -\frac{K^{\nu_{\textrm{E2}}}_{\rm FS}}{\nu_{\textrm{E2}}} \right)  \langle r_N^2 \rangle \simeq 2.4\times 10^{-3}\,.
\label{eq:factor_Yb}
\eea
For a physical insight, we give an order-of-magnitude estimate of the FS coefficient  in the Supplemental Material; which is only a rough approximation and not a substitute for the detailed calculation presented here.

We experimentally demonstrate the proposed method using a single-ion \textsuperscript{171}Yb\textsuperscript{+} optical clock~\cite{Huntemann2016,Sanner2019a}. 
A single trapped ion is probed in the E3 and E2 transitions in an alternating fashion using laser pulses with wavelengths of about 467~nm and 435~nm, respectively (see~\cite{Lange2021,Filzinger:2023zrs} for
details on the clock operation). The E3 transition is interrogated with a Ramsey dark time of 500 ms. For the E2 transition, the natural lifetime of the excited state of about 50~ms limits the interrogation time, and we typically use a single 42~ms Rabi pulse. 

The frequency ratio measurement is determined by the atomic reference for averaging intervals larger than about 200\,s. Then, the measurements of $\nu_{\textrm{E3}}/\nu_{\textrm{E2}}$ are limited by white frequency noise, given by the quantum projection noise due to the limited interrogation time of the E2 transition. The measurement instability is $1.0 \times 10^{-14} /\sqrt{\tau}$, where $\tau$ is the averaging time in seconds.

We analyze about 235 days of data taken in a total period $T$ of about 26 months and search for sinusoidal modulations as described in~\cite{Filzinger:2023zrs}. 
We find no modulation with an amplitude exceeding significantly that expected from the quantum projection noise. 
The upper 95\% confidence levels of the extracted oscillation amplitudes yields largely frequency-independent limits below about $2\times 10^{-17}$ on the relative amplitudes for frequencies in the range $1/T\approx 1.4\times10^{-8}\,$Hz to $0.005\,\textrm{Hz}$. 
For frequencies smaller than $1/T$ (corresponding to DM masses below $6.0\times 10^{-23}\,$eV), where our data cover less than a full oscillation cycle, the limits on the amplitude increase since being near an antinode of an oscillation cannot be ruled out. 

Since we did not find any statistically significant sinusoidal modulations in our data, we can use our results to constrain any model that would lead to such modulations. 
Using the relation between oscillations in the frequency ratio $\nu_{\textrm{E3}}/\nu_{\textrm{E2}}$ and the UDM couplings $1/f_a$, as well as $d_g$ and 
$d_{\hat{m}}$ given in Eq.~\eqref{eq:QCD_axion_cr} and Eq.~\eqref{eq:scalar_DM_cr} respectively, we derive limits for these couplings. 
Here, we assume that the UDM field of mass $m_{\phi}$ ($m_a$) comprises all of the DM with $\rho_{\rm DM} = 0.4 \GeV/(\rm cm)^3$. 
Note that for UDM masses below $\approx 10^{-22}\eV$, this assumption needs to be relaxed, leading to weakened bounds for these masses, which is not considered in any of the constraints plotted.
Different UDM models predict possible local over- or under-densities compared to the standard halo model, e.g. solar halo~\cite{halo}, bosenova~\cite{bosenova}, and streams~\cite{streams}. Other papers predict large density enhancement close to the surface of the earth. See \cite{Leane:2022hkk} for the case of $\sim {\cal O} (1)$ GeV DM mass, and enhancement due to gravitational effects in \cite{Prezeau:2015lxa} and \cite{Sofue:2020mda}. The present analysis is sensitive to square root of the local DM density and a detailed analysis in terms of the individual models would be required to determine limits for different densities, signal durations, and the  UDM coherence properties.

The largest DM mass included in our analysis is approximately $2\times 10^{-17}\,\textrm{eV}$, which has a coherence time of more than 6 years, well above our total measurement period of $T\approx 2\,$years. Thus, we do not need to include DM decoherence in our analysis. We take into account stochastic fluctuations of the DM amplitude and correspondingly re-scale our limits by a factor of 3~\cite{Centers2021}.
\begin{figure}[h!]
	\centering
	\includegraphics[width=\columnwidth]{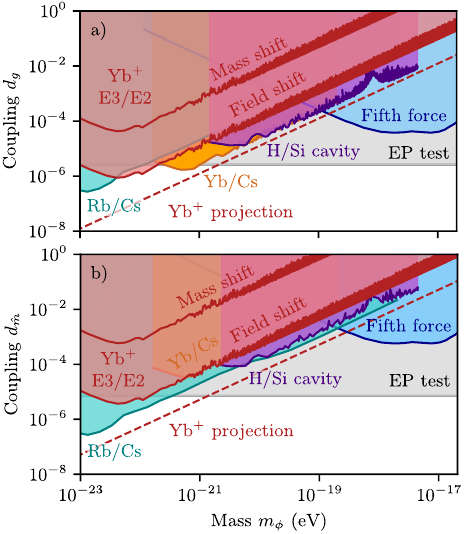}
    \caption{Exclusion plot for the linear scalar DM coupling a) to the gluons $d_g$ and b) to the quark masses $d_{\hat m}$ as a function of DM mass $m_\phi$. 
    Using the field shift effect, limits at the 95\% confidence level from long-term measurements of the frequency ratio $\nu_{\textrm{E3}}/\nu_{\textrm{E2}}$ in a single-ion optical clock are shown in dark red. 
    Based on the same experiment, the much weaker limit from the mass shift is shown for reference. The dashed line shows a projection assuming amplitude limits at the $1\times10^{-18}$-level. 
    The grey and the blue lines depict the strongest EP bound~\cite{MICROSCOPE:2022doy} and the bound from various fifth force searches~\cite{Fischbach:1996eq}, respectively. 
    Bounds from existing spectroscopy experiments are also shown: Rb/Cs~\cite{Hees:2016gop} (turquoise), Yb/Cs~\cite{Kobayashi2022} (orange), H/Si~\cite{Kennedy:2020bac} (purple).}  
	\label{fig:dg}
\end{figure}
In Fig.~\ref{fig:dg}, we show the exclusion plot of the scalar UDM coupling $d_g$ to gluons and $d_{\hat{m}}$ to the quark masses as a function of DM mass, $m_\phi$. 
Our limits are competitive compared to other spectroscopic limits~\cite{Hees:2016gop,Kobayashi2022,Kennedy:2020bac}, but importantly rely on a completely different effect, which makes our search complementary to previous results. We set new limits on the coupling $d_g$ for masses around $10^{-22}$\,eV. For reference, we also plot the much weaker limits derived from the mass shift. This effect is suppressed here, but it can be used instead of the field shift to probe the nuclear degrees of freedom with optical clocks based on light elements. While bounds from EP tests and fifth-force searches are more stringent than spectroscopic bounds for most masses within the range investigated here, we note that for a non-generic coupling of scalar UDM to the SM content, bounds from the EP-violation and fifth-force experiments may be further suppressed by a factor $\mathcal{O}(10^{-3})$~\cite{Banerjee:2022sqg,Oswald:2021vtc}.
\begin{figure}[h!]
	
	\includegraphics[width=\columnwidth]{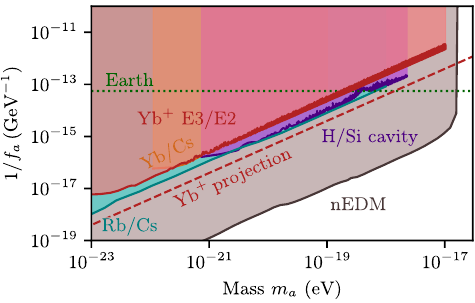}
	\caption{Exclusion plot for the QCD axion coupling $1/f_a$ as a function of the axion mass, $m_a$. The limits based on the long-term measurements of the frequency ratio $\nu_{\textrm{E3}}/\nu_{\textrm{E2}}$ in a single-ion optical clock are shown in dark red. The dashed line is a projection assuming amplitude limits at the $1\times10^{-18}$-level. Existing limits based on oscillating neutron electric dipole moment~\cite{Abel:2017rtm} are shown in brown, and theory limits due to density effects of the Earth~\cite{Hook:2017psm} as a dotted green line. Bounds from existing spectroscopy experiments are also shown: Rb/Cs~\cite{Hees:2016gop} (turquoise), Yb/Cs~\cite{Kobayashi2022} (orange), H/Si~\cite{Kennedy:2020bac} (purple). 
    }
	\label{fig:fa}		
\end{figure}
In Fig.\,\ref{fig:fa}, we show the parameter space of axion-gluon coupling of Eq.\,\eqref{eq:QCD_axion_cr} as a function of the axion mass, $m_a$. Our limits do not currently exceed those of experiments that search for an oscillating neutron electric dipole moment,~\cite{Abel:2017rtm}.  
However, future investigations using dynamical decoupling techniques~\cite{Aharony:2019iad,Kennedy2020} can extend the search towards higher masses into a previously experimentally unexplored regime. In this context, we note that the bound associated with Earth~\cite{Hook:2017psm} is not related to a search for oscillating energy levels. It originates from the fact that for small enough $f_a$ the Earth's matter density affects the axion potential, driving it away from zero. This bound can possibly be avoided if one introduces a new interaction between the axion and the SM matter fields. In the forthcoming work, the analysis in this paper will also be extended to higher frequencies (up to $\sim$100 MHz) based on the experimental data from atomic \cite{Tretiak2022} and molecular \cite{OswaldPRL2022} spectroscopy.

The measurement can be improved by accumulating more data or, given a certain measurement time, improving its instability. Since the $\nu_{\textrm{E3}}/\nu_{\textrm{E2}}$ measurement instability is limited by the finite lifetime of the $E2$ excited state, comparing the $E3$ clock to a clock with superior stability and suitable sensitivity will lead to an improved search. The projections shown in the plots assume amplitude limits at the level $1\times 10^{-18}$, which could be obtained with the present $\nu_{\textrm{E3}}$ instability and similar measurement time.

In summary, we show that UDM interacting with the QCD sector leads to oscillations of the nuclear charge radius and consequently of electronic transition frequencies, which can be investigated with high precision in optical clocks. We apply this idea to two transitions in \textsuperscript{171}Yb\textsuperscript{+}. A long-term measurement of the frequency ratio, and the calculated sensitivities, provide constraints on the coupling of UDM to quarks and gluons. 
While these results only improve the coupling $d_g$ for a small mass range, they constitute, to our knowledge, the first investigation of UDM-nuclear couplings using an optical atomic clock comparison. Future investigations based on the derived principle, employing combinations of optical clocks promising larger sensitivity, in particular, those based on highly charged ions \cite{Kozlov2018,Rehbehn2021}, are expected to investigate couplings well below the current parameter range.

\section*{Acknowledgements}
We would like to thank Hyungjin Kim, Eric Madge, Ziv Meir, Gerald Miller, Shmuel Nussinov, Roee Ozeri, Ekkehard Peik, and Antonio Pineda for useful discussions. The work of AB is supported by the Azrieli foundation. The work of DB is supported in part by the Deutsche Forschungsgemeinschaft (DFG) - Project ID 423116110 and Cluster of Excellence ``Precision Physics, Fundamental Interactions, and Structure of Matter'' (PRISMA+ EXC 2118/1) funded by the DFG within the German Excellence Strategy (Project ID 39083149). 
The work of GP is supported by grants from BSF-NSF, Friedrich Wilhelm Bessel research award, GIF, ISF, Minerva, SABRA Yeda-Sela WRC Program, the Estate of Emile Mimran, and the Maurice and Vivienne Wohl Endowment. 
The work of MS was supported in part by the NSF QLCI Award OMA - 2016244, NSF Grants PHY-2012068, and PHY-2309254. The work of MS and SP was supported by the European Research Council (ERC) under the European Union’s Horizon 2020 research and innovation program (Grant Number 856415).
The work of MF and NH was supported by the 
DFG under SFB~1227 DQ-\textit{mat} -- Project-ID 274200144 -- within project B02 and the Max Planck--RIKEN--PTB Center for Time, Constants and Fundamental Symmetries.

\appendix

\section*{Supplemental Material}

\subsection{Calculation of $\alpha$ and $\beta$}

We want to estimate how the nuclear charge radius depends on the fundamental parameters $\Lambda_{\rm QCD}$ and $\theta_{\rm eff}$. Since the nuclear charge radius is expected to scale as $A^{1/3}$, where $A$ is the atomic mass number, the contribution of the charge radii of individual nucleons is expected to be small for heavy nuclei. Even for a light nucleus such as the deuteron, the contribution of the iso-scalar charge radius of the nucleon is smaller than the next-to-leading order contribution in the pion-less effective field theory~\cite{Chen:1999tn}. We therefore expect that the dominant contribution comes from the nuclear size itself.  As explained in the main text, to estimate it, we consider a relativistic mean field approach to model the nuclear interactions (see for instance~\cite{Niksic:2011sg,Ring:1996qi,Reinhard:1989zi} and Refs. therein). 
The nuclear mean field is described by a spin-half field, $\Psi$, and the most important interaction term is mediated by the lowest lying isospin-singlet scalar meson ($\phi$)~\cite{Walecka:1974qa}. The Lagrangian can be written as,
\bea
\!\!\!\!\!
\mathcal{L} \supset \bar\Psi\left(i\slashed{\partial}-m_N\right)\Psi+ \frac{1}{2}\left(\partial_\mu\phi\right)^2-\frac{1}{2}m_s^2\phi^2 -g_s\phi \bar\Psi \Psi\,,
\eea
where $g_s$ is the coupling strength, and $m_s$ and $m_N$ are the mass of the scalar and nucleons respectively. 
In the context of the mean field theory along with a scalar one can also consider a vector meson mediating interaction between the nucleons  (see {\it e.g.}~\cite{Horowitz:1981xw} and refs. therein). As we are interested in only how the charge radius of the nucleus depends on fundamental parameters such as $\Lambda_{\rm QCD}$ and $\theta_{\rm eff}$, we omit such contributions to simplify the discussion.
We expect that neglecting the vector meson mediated interaction between the nucleons essentially means neglecting terms of the order of $\mathcal{O}(m_s^2/m_V^2)$ 
in the nuclear potential, where $m_V $ is the mass of the vector meson. 
Also, as we are interested in heavy nuclei with a large number of nucleons, the contribution to the nuclear force  from the spin-dependent one-pion exchange averages to zero~\cite{Miller:1972zza,Brockmann:1978zz,Horowitz:1981xw}.
\\

In the main text, we parameterize the dependence respectively as,  
\bea
\alpha= \frac{\partial\ln \left<r^2\right>}{\partial\ln\Lambda_{\rm QCD}},\,\,{\rm and}\,\,\beta= \frac{\partial\ln \left<r^2\right>}{\partial\ln m_\pi^2}\,,
\label{eq:alpha_beta_supp_mat}
\eea
where, $\left<r^2\right>$ is the charge radius of the nucleus. Below we will see that the $\theta_{\rm eff}$ dependence arises via the dependence of $\left<r^2\right>$ on $m_\pi^2$.  
In the following we calculate the exact dependence of $\left<r^2\right>$ on $g_s,\,m_s$, and $m_N$, but one should keep in mind that to calculate  $\alpha$ and $\beta$ we only need the power dependence on $g_s,\,m_s$, and $m_N$. Multiplicative parameters will cancel in the logarithmic derivative.

The equations of motion (EOM) of $\Psi$ and $\phi$ are
\bea
\left(\Box+m_s^2\right)\phi &=& -g_s \bar\Psi\Psi\,,\label{eq:phi_eom}\\
\left[i\slashed\partial-m_N-g_s\phi\right]\Psi&=&0\,.
\label{eq:psi_eom}
\eea
To solve this set of coupled equations, numerically iterative methods are usually used in the literature~\cite{Horowitz:1981xw,Furnstahl:1987rd}. 
As we are interested in the parametric dependence of the nuclear size on fundamental parameters, we want to obtain an analytic expression. To do that, we follow the process outlined in~\cite{Horowitz:1981xw}. 

Following ~\cite{Neubert:1993mb,Manohar:1997qy} we obtain the non-relativistic (NR) limit of the fermion EOM for $\psi$, the NR limit of $\Psi$. It takes the form of 
the Schr\"odinger equation as
\bea
\left(-\frac{\nabla^2}{2m_N}+V(r)\right)\psi(\vec r,t) = i\frac{\partial}{\partial t}\psi(\vec r,t)\,,
\label{eq:fermion_NR}
\eea
with,
\bea
V(r)= g_s\phi(r)\,. 
\eea
Thus we obtain, perhaps as expected, that the scalar field background acts as a potential for the fermions. 
The energy eigenvalues $E$ are obtained from
\bea
\left[-\frac{\nabla^2}{2m_N}+V(r)-E\right]\psi(\vec r) =0.
\label{Eq:NR_SE}
\eea

As we are interested in obtaining the radius of a nucleus, using Eq.~\eqref{eq:phi_eom}, we approximate $\phi(r)$ as
\bea
\phi(r)\approx \phi_0(r) =- \frac{g_s \bar\Psi\Psi}{m_s^2}=-\frac{g_s\, \psi^\dagger(r)\psi (r)}{m_s^2}\,,
\label{eq:phi_bg}
\eea
using that at the lowest order in $1/m_N$, $\bar\Psi\Psi=\bar\psi\psi=\psi^\dagger\psi$. 
The above approximation is valid as long as $m_s^2\phi \gg \Box^2\phi$ which parametrically becomes, $m_s a \gg 1$ for some characteristic length scale $a$.

Now, equipped with all these, 
we want to solve Eq.~\eqref{Eq:NR_SE}. 
As done numerically in~\cite{Horowitz:1981xw}, we start with the zeroth-order assumption that the potential, $V(r)$, that describes a bound nucleus is only non-zero in some characteristic length scale $a$. 
Also in that range, it is negative and constant. 
Thus, at the zeroth order, we replace $V(r)$ in Eq. (\ref{Eq:NR_SE}) by $V^{(0)}(r) = V_0 \theta(a-r)$, where $V_0=-g_s^2/(m_s^2a^3)$. 
We now find for $\psi_0$ the bound state solution, i.e., $V_0<E<0$, of the well-known problem of a spherical potential well. Defining $k_3=\sqrt{-2m_N E}$, and $k_2 = \sqrt{2 m_N(V_0-E)}$, the bound state energies are found by solving a transcendental equation 
\be \label{energy_condition}
k_2\coth{k_2a}+k_3=0
\ee 
Using this solution we find at the zeroth-order
\bea
\left<r^2\right>_0 = \frac{\int d^3r\, r^2\, |\psi_0(r)|^2}{\int d^3r\,|\psi_0(r)|^2}= a^2 \times f\,,
\eea
where $f$ is an order-one function of $2m_NV_0a^2$.  

As discussed before, for a heavy nucleus, we expect the dominant contribution to the charge radius to come from the nuclear size itself, and thus the zeroth-order charge radius is dictated by the input parameter $a$ multiplied by some $\mathcal{O}(1)$ number which depends on the nuclear parameters above. However, as $a$ being the input parameter, one can always redefine it to match the radius at zeroth order. 
Thus at this iteration, the charge radius does not have dependence on $g_s$, $m_s$ and/or $m_N$.

To obtain the charge radius dependence on those parameters, we go
beyond the step-function approximation of the nuclear potential. Let $\psi_0(r)$ be the solution to Eq.~\eqref{Eq:NR_SE} with a step function potential, namely with $V(r)=V^{(0)}(r)$. We would like to find $\psi_(r)$ that is the solution of 
\bea
 &&\left[-\frac{\nabla^2}{2m_N}-\frac{g_s^2}{m_s^2}(\psi_0^\dagger\psi_0)-E\right]\psi(\vec r) = 0.
\label{eq:fermion_NR}
\eea
Adding and subtracting $V^{(0)}(r)$ we get
\bea
&&\left[-\frac{\nabla^2}{2m_N}+V^{(0)}(r)-E\right]\psi(\vec r) = \delta V(r) \psi(\vec r)\,,
\label{eq:deltaV}
\eea
where $\delta V(r) =V^{(0)}(r)+g_s^2(\psi_0^\dagger\psi_0)/m_s^2$. The above equation can be solved by the Green's function method by identifying the right hand side as the source term of a homogeneous equation. 
The Green's function, $G(\vec{r},\vec{r}^{\,\prime},E)$, satisfies the equation 
\be
\left[-\nabla^2/(2m_N)+V^{(0)}(r)-E\right]G(\vec{r},\vec{r}^{\,\prime},E)=\delta(\vec{r}-\vec{r}^{\,\prime})\,,
\ee
Define $\tilde{G}$ as the ``reduced" Green's function~\cite{Friar:1978wv} obtained from $G(\vec{r},\vec{r}^{\,\prime},E)$ by subtracting $\psi_0(r)\psi_0^*(r')/(E_0-E)$.  The solution of Eq.~\eqref{eq:deltaV} is $\psi(\vec r)=\psi_0(r)+\psi_1(r)$, where 
\bea
\psi_1(r)= \int d^3r'\, \tilde{G}(\vec{r},\vec{r}^{\,\prime},E)\,\delta V(r')\,\psi_0(r')\,.
\eea

We now decompose the Green's function to partial waves and consider only its $L=0$ part. Transforming to the effective 1-D problem in the standard way, we need the solution of  
\bea
\!\!\!\!\!\!\!\!
\left[-\frac1{2m}\frac{\partial^2}{\partial r^2}+V^{(0)}(r)-E\right]G(r,r^\prime,E)=\delta(r-r^\prime)\,.
\eea

The solution for such an equation can be found by following the procedure of ~\cite{Baltin:1985}, keeping in mind that the Green's function of ~\cite{Baltin:1985} is the solution of $(E-H)G=\delta(r-r^\prime)$. We refer the readers to~\cite{toappear} for the complete analysis. 
Using it we calculate the contribution to the charge radius due to $\psi_1(r)$, namely, $\left<r^2\right>=\left<r^2\right>_0+\left<r^2\right>_1$, where
\begin{align}
  \left<r^2\right>_1 \!= 2\! \!\int\! d^3s d^3s' \left[s^2-a^2\right] \psi_0(s')\tilde{G}(s,s^{\,\prime},E)\delta V(s')\psi_0(s)\,,
  \label{eq:r1}
\end{align}
and we set $\left<r^2\right>_0\equiv a^2$.
Thus the total charge radius of the nucleus can be written as,
\bea \label{R01_defined}
\left<r^2\right>= a^2(1+ R_{01})\,,
\eea
where $R_{01}\equiv \left<r^2\right>_1/a^2$. 

Consider the dependence of $\left<r^2\right>$ on a parameter $p$. We have from Eq.~\eqref{R01_defined} 
\bea \label{logp}
\frac{\partial\ln \left<r^2\right>}{\partial\ln p}=\frac{R_{01}}{1+R_{01}}\frac{\partial\ln R_{01}}{\partial\ln p}.
\eea

Now we want to obtain the parametric dependence of $R_{01}$ on $g_s$, $m_s$ and/or $m_N$. Define $k_2\equiv i k$. Using the bound state energy condition, Eq.~\eqref{energy_condition}, and $(ak_2)^2-(ak_3)^2=2 m_N V_0 a^2$, we have
\be\label{variable_change}
-\frac{2 g^2_s\, m_N}{m_s^2\, a}=2m_N V_0a^2=- \frac{(k\,a)^2}{\sin(k\,a)^2}\,.
\ee
Defining the dimensionless variables  $x=s/a$, $x'=s'/a$ in Eq.~\eqref{eq:r1} and using Eq.~\eqref{variable_change}
we obtain
\bea
R_{01} =  - \frac{(k\,a)^2}{\sin(k\,a)^2} \times \I(k a)\,.
\label{eq:R01_exp}
 \eea
The dimensionless integral $\I(k a)$ is made of six different integration regions depending on the values of $x$ and $x^\prime$. 
In~\cite{toappear} we explicitly calculate these integrals. 
Thus Eq.~\eqref{eq:R01_exp} gives us the parametric dependence of the charge radius on the fundamental theory parameters.

The ground state energy is obtained for $\pi/2<k a<\pi$. 
For the ground state we find that the pre-factor in Eq.~\eqref{logp} is \cite{toappear}
\bea \label{prefactor}
\frac{R_{01}}{1+R_{01}} =0.995\pm 0.005\,, 
\eea
where the uncertainty is obtained by varying $\pi/2<k a<\pi$.

To calculate $\alpha$, we note that 
$\partial \ln m_s/\partial\ln\Lambda_{\rm QCD}=1$~\cite{Zwicky:2023krx,Hashimoto:2021ihd,Pelaez:2006nj} (from both the large $N_c$-chiral and holographic approaches), and $\partial \ln m_N/\partial\ln\Lambda_{\rm QCD}=0.9$~\cite{OswaldPRL2022,Shifman:1978zn,Hill:2016bjv}. 
Using Eq.~\eqref{eq:alpha_beta_supp_mat}, Eq.~\eqref{logp} and Eq.~\eqref{eq:R01_exp}, we obtain
\bea
\alpha = \frac{\partial\ln \left<r^2\right>}{\partial\ln\Lambda_{\rm QCD}}\simeq -1.1\,,
\eea
with an uncertainty of $\pm 0.006$ from Eq.~\eqref{prefactor}, with additional sources of uncertainty discussed below.

Similarly, to obtain $\beta$, we note the pion mass dependence of the theory parameters:   $\partial \ln m_N/\partial\ln m_\pi^2\simeq 0.06$~\cite{Kim:2022ype}, and $\partial \ln (g_s^2/m_s^2)/\partial\ln m_\pi^2\simeq -0.4$~\cite{Ubaldi:2008nf,Lee:2020tmi}. 
We obtain from Eq.~\eqref{eq:alpha_beta_supp_mat}, Eq.~\eqref{logp} and Eq.~\eqref{eq:R01_exp},  
\bea
\beta = \frac{\partial\ln \left<r^2\right>}{\partial\ln m_\pi^2} \simeq -0.34\,.
\eea 
with an uncertainty of $\pm 0.002$ from Eq.~\eqref{prefactor}, with additional sources of uncertainty discussed below.

As we derive an analytical estimate for $\alpha$ and $\beta$ rather than relying on numerical methods, we also assess the associated theoretical error bars. 
Beyond the quoted uncertainties, our estimate is based on a mean-field description of the nucleus, which becomes exact only in the $A \to \infty$ limit. 
For ${}^{171}\text{Yb}^+$, this introduces a theoretical error of approximately $0.6\%$. 
Additionally, as discussed earlier, we consider only the nuclear potential arising from scalar meson exchange between nucleons, neglecting contributions from vector meson-mediated interactions, and possibly through other heavier meson exchanges. This approximation results in an additional $30\%$ uncertainty due to neglecting terms of the order $\mathcal{O}(m_s^2/m_V^2)$, where we take $m_s = m_\sigma = 443 \MeV$~\cite{Acharya:2015pya} and $m_V = m_\rho = 770 \MeV$~\cite{ParticleDataGroup:2020ssz}.

\subsection{Order of magnitude estimate of $K_{\rm FS}$}

We can obtain an order of magnitude estimate of $K_{\rm FS}$ by using quantum mechanical first order perturbation theory and the fact that the nuclear size is small compared to the atomic size. The energy level shift is~\cite{Friar:1978wv}
\bea
(\Delta E)_{\rm FS}=\frac{2\pi}{3}\left|\psi(0)\right|^2Z\alpha\left<r_N^2\right>,
\eea
where $\psi$ is the wave function of the state, and $\alpha$ denotes the fine structure constant here. 
For an $s$-wave of a valence electron in a heavy, neutral, multi-electron atom, we get  $\left|\psi_s(0)\right|^2=Z/a_0^3$~\cite{Budker:2008},
where $a_0\approx 0.53\cdot 10^5$ fm is the Bohr radius. In $\hbar=1, c=1$ units,  $1 \mbox{ fm}^{-1}=3\cdot10^{14}\mbox{ GHz}$. For Yb $Z=70$, which gives an energy shift of  
\bea
\Delta E\approx\frac{2\pi}{3}Z^2\alpha\frac{\left<r_N^2\right>}{a_0^3}\approx150 \frac{\mbox{GHz}}{\mbox{fm}^2}\left<r_N^2\right>\,.
\label{eq:estimate_FS}
\eea
Up to a sign and within an order of magnitude this agrees with the calculated values of $K^\nu_{\rm FS}$. 
We emphasize that this is only an order of magnitude estimate and it does not replace the detailed calculation in the main text.

\bibliographystyle{apsrev4-1}
\bibliography{charge_radius_ref_v4}

\end{document}